\documentclass[twocolumn]{aastex61}

\usepackage{graphicx}
\usepackage{amsmath}
\usepackage{amssymb,latexsym,amsopn}
\usepackage{color}

\usepackage{ulem}

\shorttitle{MAXI J1957$+$032}
\shortauthors{Ravi}

\begin{document}

\title{MAXI J1957$+$032: an accreting neutron star possibly in a triple system}

\author{V. Ravi}
\affil{Cahill Center for Astronomy and Astrophysics, MC 249-17, California Institute of Technology, Pasadena, CA 91125, USA. E-mail: vikram@caltech.edu}

\begin{abstract}

I present an optical characterization of the Galactic X-ray transient source MAXI\,J1957$+$032. This system flares by a factor of $\gtrsim10^{4}$ every 
few-hundred days, with each flare lasting $\sim5$~days. I identify its quiescent counterpart to be a late-K/early-M dwarf star at a distance of $5\pm2$\,kpc. 
This implies that the peak $0.5-10$\,keV luminosity of the system is $10^{36.4\pm0.4}$\,erg\,s$^{-1}$. As found by Mata Sanchez et al., the outburst properties of MAXI\,J1957$+$032 are most consistent with the sample of accreting millisecond pulsars. However, the low inferred accretion rate, and the lack of evidence for a 
hydrogen-rich accretion flow, are difficult to reconcile with the late-K/early-M dwarf counterpart being the mass donor. Instead, the observations are best described 
by a low-mass hydrogen- and possibly helium-poor mass donor, such as a carbon-oxygen white dwarf, forming a tight interacting binary with a neutron star. The observed main-sequence counterpart would then likely be in a wide orbit around the inner binary. 

\end{abstract}


\section{Introduction}

The {\it Monitor of the All-sky X-ray Image} (MAXI) aboard the International Space Station scans 
95\% of the sky at energies between $0.4-30$\,keV during each 92\,min orbit \citep{mku+09}. The Gas-Slit Camera (GSC), operating between $2-30$\,keV, 
is the more sensitive of the two MAXI instruments, capable of detecting a $\gtrsim10$\,mCrab source and localizing it to $3$\,deg$^{2}$ in a single day of observations. 
One of the primary science goals of the MAXI/GSC instrument is realized through its nova alert system \citep{nmo+10}, which 
automatically identifies outbursts in the day-averaged data and provides alerts to the interested community within tens of seconds. 

Transient X-ray emission was first identified from a new source, MAXI\,J1957+032 (hereafter J1957), by the MAXI/GSC nova alert 
system on 2015 May 11 \citep[ATel 7504;][]{nsm+15}, with a peak flux in the 2$-$10\,keV energy band of $10$\,mCrab. This flare was simultaneously detected by the IBIS/ISGRI telescope aboard the INTEGRAL mission \citep[ATel 7506;][]{cml+15}, with a flux of 
4.1(7)\,mCrab in the 20$-$60\,keV band.\footnote{Throughout this paper, $1\sigma$ uncertainties in the last significant figures are provided in 
parentheses.} The MAXI flare lasted for two days. A re-brightening of the source was detected by 
MAXI/GSC on 2015 October 6 \citep[ATel 8143;][]{sns+15}, and a peak flux of 16(5)\,mCrab was reached on 2015 October 7 in the 
2$-$20\,keV band. Again, the flare lasted for two days in the MAXI energy band. Two further flares were detected by 
MAXI on 2016 January 7 \citep[ATel 8529;][]{tnu+16} and 2016 September 29 \citep[ATel 9565;][]{nst+16}, with comparable fluxes and durations. 

J1957 has so far eluded unambiguous characterization. Triggered observations with the {\it Swift} X-Ray Telescope \citep[XRT;][]{bhn+05} on 2015 May 13, 
following the first reported MAXI flare, detected a single source within the MAXI/GSC error circle that faded by a factor of $\sim100$ 
over the following week \citep[ATels~7506, 7520;][]{cml+15,mlp+15}. A blue optical counterpart coincident with the 
{\it Swift}/XRT $3$\arcsec~localization was detected by the GROND camera at the MPI/ESO 2.2\,m telescope on 2015 May 15, 
with magnitude $r'=20.0(1)$, and fading by 1.2 magnitudes by the following night \citep[ATel~7524;][]{ryg15}. The quiescent 
optical counterpart was recently identified, with magnitude $R=21.4(2)$, by the SALTICAM imager at the South African Large Telescope (SALT) \citep[ATel~9649;][]{bkc+16}. 
A  SALT spectrum of J1957 was obtained following the 2016 September 29 outburst, although only a featureless blue continuum was detected \citep{bkc+16,mca+17}. 
\citet{mca+17} interpreted the spectral properties of J1957 during outburst as corresponding to an X-ray irradiated 
disk, which when considered together with the X-ray outburst lightcurves were thought to be suggestive of an accreting millisecond 
X-ray pulsar (AMXP).

I entered the fray following the second reported outburst of J1957 in October 2015, by collating what was known about this source, 
and attempting to identify the quiescent optical counterpart using deep observations at the W.~M.~Keck Observatory. On 2016 September 30, I 
was also able to obtain a spectrum of the optical counterpart to the fourth reported MAXI outburst. A summary and re-analysis 
of the X-ray observations is presented in \S2, and the optical observations are presented in \S3. 

The properties of J1957, including the outburst and quiescence timescales, brightness, and detection of optical flares, 
are most characteristic of a Galactic X-ray binary system. That is, the outbursts are most likely driven by accretion onto a compact 
object from a companion star. This paper therefore attempts to answer the following, in pedagogical order:
\begin{enumerate}
\item What is the nature of the quiescent optical counterpart?
\item Is the compact accretor a white dwarf (WD), neutron star (NS), or black hole (BH)?
\item How does the accretion occur?
\item What is the evolutionary state of the system?
\end{enumerate}
I present the interpretation in \S4, and conclude in \S5.

\section{MAXI J1957$+$032: X-ray observations}

\begin{deluxetable}{ccc}
\tabletypesize{\scriptsize}
\tablecaption{Summary of MAXI/GSC detections of outbursts from MAXI J1957$+$032.}
\tablewidth{0pt}
\tablehead{
\colhead{Epoch (MJD)} & \colhead{$\tau_{f}$ (days)} & \colhead{References}
}
\startdata
57153 & 2  & 1,2,3,4 \\
57302 (149)* & 2  & 5,6,7,8 \\
57394 (93)* & 3  & 9 \\
57660 (266)* & 2 & 10 \\
\enddata
\label{table:1}
\tablecomments{1$-$ ATel 7504; 2$-$ ATel 7506; 3$-$ ATel 7520; 4$-$ ATel 7524; 5$-$ ATel 8143; 6$-$ ATel 8146; 7$-$ ATel 8149; 8$-$ ATel 8197; 9$-$ ATel 8529; 10$-$ ATel 9565. *The quantities in parentheses are the times since the previous flare.}
\end{deluxetable}

\begin{deluxetable*}{ccccc}
\tabletypesize{\scriptsize}
\tablecaption{{\it Swift}/XRT photon-counter observations of MAXI J1957$+$032.}
\tablewidth{0pt}
\tablehead{
\colhead{Epoch (MJD)} & \colhead{ObsID} & \colhead{XRT exposure (s)} & \colhead{0.5$-$10\,keV count rate ($s^{-1}$)} & \colhead{0.5$-$10\,keV $unabs$ flux (erg\,cm$^{-2}$\,s$^{-1}$)} 
}
\startdata
57156.05 & 00033770001 & 2982 & $0.50(1)$ & $1.65(3)\times10^{-11}$ \\
57157.68 & 00033770002	& 1990 & $0.013(3)$ & $4.2(9)\times10^{-13}$ \\
57162.07 & 00033770005 & 2783 & $0.006(2)$ & $2.1(6)\times10^{-13}$ \\
57164.78 & 00033770006 & 867 & $0.008(3)$ & $3(1)\times10^{-13}$ \\
57165.80 & 00033770007 & 2164 & $0.021(3)$ & $7(1)\times10^{-13}$ \\
57193.04 & 00033770008 & 4982 & $0.006(1)$ & $2.0(5)\times10^{-13}$ \\
57304.69 & 00033770009 & 994 & $0.56(3)$ & $1.85(9)\times10^{-11}$ \\ 
57305.63 & 00033770010 & 995 & $0.055(9)$ & $1.8(3)\times10^{-12}$ \\
57306.69 & 00033770011 & 996 & $<0.0385$ & $<1.28\times10^{-12}$ \\
57307.72 & 00033770012 & 995 & $<0.0127$ & $<4.22\times10^{-13}$ \\
57308.65 & 00033770013 & 1019 & $<0.0384$ & $<1.28\times10^{-12}$ \\
57309.28 & 00033770014 & 1043 & $0.004(2)$ & $1.2(7)\times10^{-13}$ \\
57310.55 & 00033770015 & 754 & $<0.0354$ & $<1.18\times10^{-12}$ \\
57311.45 & 00033770016 & 1023 & $<0.0372$ & $<1.24\times10^{-12}$ \\ 
57662.65 & 00033770019 & 392 & $2.42(9)$ & $8.0(3)\times10^{-11}$ \\
57663.37 & 00033770020 & 1773 & $0.84(3)$ & $2.81(8)\times10^{-11}$ \\ 
57664.20 & 00033770021 & 1739 & $0.103(8)$ & $3.4(3)\times10^{-12}$ \\ 
57665.62 & 00033770022 & 1402 & $<0.0274$ & $<9.10\times10^{-13}$ \\ 
57667.54 & 00033770023 & 639 & $<0.0175$ & $<5.81\times10^{-13}$ \\ 
57668.11 & 00033770024 & 2040 & $<0.0073$ & $<2.42\times10^{-13}$ \\ 
57669.79 & 00033770025 & 1941 & $<0.0201$ & $<6.68\times10^{-13}$ \\ 
57731.27 & 00033770026 & 8185 & $<0.0023$ & $<7.6\times10^{-14}$ \\
\enddata
\label{table:xrt}
\tablecomments{The epoch MJDs are the mid-points of the {\em Swift}/XRT exposures. The count rates are corrected for the background, vignetting, exposure, and the point-spread function weighted by the exposure maps. The unabsorbed ($unabs$) fluxes are calculated assuming a photon index of $\Gamma=1.92$, and an absorbing column of $N_{\rm H}=1.7\times10^{21}$\,cm$^{-2}$. All upper limits are at the $3\sigma$ confidence level.}
\end{deluxetable*}

The MAXI/GSC nova alert system has triggered on four outbursts from J1957 to date, as summarized in Table~\ref{table:1}. I also list the outburst timescales, and relevant references for observations of each outburst. I note that the publicly available MAXI/GSC lightcurves for the position of J1957 include data since MJD\,55058.  I encourage the MAXI team to analyze their data on J1957 prior to the first reported flare on MJD\,57153, to identify any other outbursts that may not have triggered the nova alert system. 

The {\it Swift} satellite conducted 28 observations of the position of J1957, in response to the MJD\,57153, MJD\,57302, and MJD\,57660 outbursts. These observations were analyzed by \citet{mca+17}. Following each outburst, rapidly fading X-ray emission was detected with the XRT, and rapidly fading UV/optical emission was detected with the 
UV/Optical Telescope \citep[UVOT;][]{rkm+05}. The maximum detected flux, observed following the MJD\,57660 outburst, was $9.1(1)\times10^{-10}$\,erg\,cm$^{-2}$\,s$^{-1}$. \citet{mca+17} fitted an absorbed power-law model to the XRT data, finding a largely consistent photon index of $\Gamma\approx2$, and an absorbing column density of $n_{H}=1.7\times10^{21}$\,cm$^{-2}$ \citep[see also ATel~9591;][]{cjm16}. No periodicities were detected in the X-ray lightcurves by \citet{mca+17}. By stacking the XRT non-detections, \citet{mca+17} placed an upper limit on the quiescent X-ray flux of J1957 of $1-3\times10^{-13}$\,erg\,cm$^{-2}$\,s$^{-1}$ (95\% confidence). 

\begin{figure}
\centering
\includegraphics[angle=-90,scale=0.47]{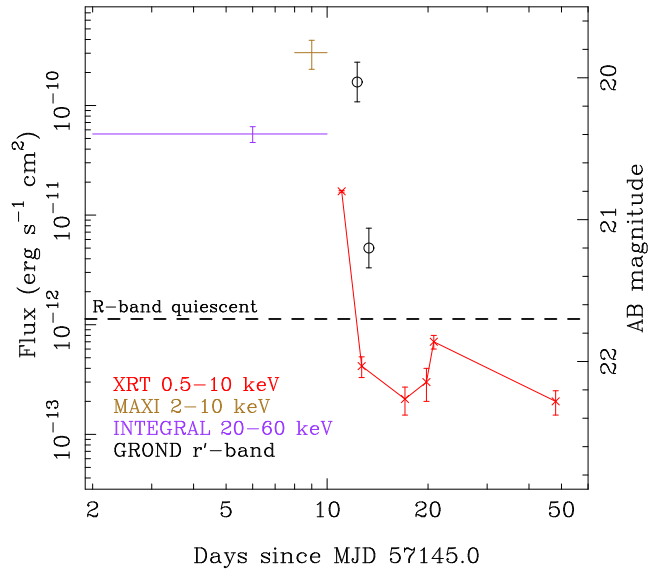}
\caption{X-ray and optical data on the MJD\,57153 flare of MAXI~J1957$+$032, reported by MAXI/GSC on 2015 May 11. The red crosses joined by a curve show the $0.5-10$\,keV {\em Swift}/XRT detections from Table~\ref{table:xrt} pertinent to this flare. The purple and gold points indicate the times and fluxes of the INTEGRAL $20-60$\,keV \citep[ATel 7506;][]{cml+15} and MAXI $2-10$\,keV \citep[ATel 7504;][]{nsm+15} detections respectively. All X-ray fluxes are corrected for absorption. The black open circles show the 
GROND $r'$-band measurements \citep[ATel~7524;][]{ryg15}, and the dashed line corresponds to the quiescent R-band magnitude measured in the Epoch 5 Keck observations discussed below (Table~\ref{table:3}). The X-ray flux scale is on the left-hand ordinate, and the optical flux scale is on the right-hand ordinate.}
\label{fig:lc}
\end{figure}

I re-analyzed the {\it Swift}/XRT data, and confirmed most of the results of \citet{mca+17}. The one exception is that I detect J1957 with a flux of $\sim3\times10^{-13}$\,erg\,cm$^{-2}$\,s$^{-1}$ up to 40 days following the MJD\,57153 outburst. In Table~\ref{table:xrt}, I present my re-analysis of the XRT observations targeted at J1957 and obtained in the photon-counting mode. Using the HEASOFT distribution v6.21, each observation was reduced using the {\em xrtpipeline} task with standard inputs. Images in the 0.5$-$10\,keV band (XRT channels 50$-$1000) were then extracted using the {\em xselect} task. I then used the {\em sosta} command of the {\em ximage} task to estimate the count rate at the position of J1957. The resulting source count rates were extracted and background-corrected using box sizes that maximized the signal to noise ratio, and were additionally corrected for vignetting, exposure, and the point-spread function (PSF) weighted by the {\em xrtpipeline}-generated exposure maps. For observations where no significant emission was detected at the position of J1957, I report $3\sigma$ upper limits. Finally, all count rates were converted to unabsorbed 0.5$-$10\,keV fluxes assuming an absorbing column of $N_{\rm H}=1.7\times10^{21}$\,cm$^{-2}$, and a photon index of $\Gamma=1.92$ determined by the weighted average of the measurements of \citet{mca+17}. I illustrate the multi-wavelength lightcurve of the MJD\,57153 flare of J1957 in Figure~\ref{fig:lc}.

The cause of the discrepancy between the present work and the non-detection of J1957 between MJD\,57157$-$57193 by \citet{mca+17} is unclear. I additionally find a tentative detection of J1957 on MJD\,57309.28 ({\em sosta} false alarm probability assuming Poisson statistics of $2.65\times10^{-4}$). It is possible that the low signal-to-noise ratios of these detections were below the threshold set by \citet{mca+17}. Additionally, by stacking the weak detections with non-detections, \citet{mca+17} may have actually reduced the signal-to-noise ratio of the faint emission from J1957 detected on some epochs.

\section{The optical counterpart}

I now present observations of the optical counterpart of J1957 at the W.~M.~Keck Observatory; the observations are detailed in Table~\ref{table:3}. 
I obtained two epochs of imaging of the quiescent 
source with the Low-Resolution Imaging Spectrometer \citep[LRIS;][]{occ+95} on the Keck~1 telescope, and four epochs of 
spectroscopy with LRIS and the DEep Imaging Multi-Object Spectrograph \citep[DEIMOS;][]{fpk+03}, three of the quiescent 
counterpart, and one during its active state. The counterpart was detected on all epochs; here I focus on data from Epochs 3--6 from Table~\ref{table:3}, 
because they represent the most sensitive observations.

\begin{deluxetable*}{cccccccccc}
\tabletypesize{\scriptsize}
\tablecaption{Log of optical observations.}
\tablewidth{0pt}
\tablehead{
\colhead{Epoch} & State$^{a}$ & \colhead{Date (MJD)} & \colhead{Instrument} & \colhead{Filter$^{b}$} & \colhead{Exposures$^{c}$ (s)} & \colhead{Slitmask} & \colhead{Slit PA (deg)}  & \colhead{Grating/grism} & \colhead{Magnitude}
}
\startdata
1 & Q & 57308.265 & Keck1/LRIS & D560  & 600; 600 & 1\arcsec$\times$175\arcsec & $-45.0$$^{d}$ & 400/8500 \& 400/3400 & --  \\
2 & Q & 57362.217 & Keck1/LRIS & g & 120, 160 & -- & -- & -- & 23.5(2)$^{e}$ \\
2 & Q & 57362.217 & Keck1/LRIS & R & 120, 120 & -- & -- & -- & 22.3(2)$^{e}$ \\
3 & Q & 57519.576 & Keck1/LRIS & D560 & $2\times1200$; $4\times560$ & 1\arcsec$\times$175\arcsec & 124.5  & 400/8500 \& 400/3400 & R=22.7(5)$^{e}$  \\
4 & Q & 57543.512 & Keck2/DEIMOS & GG455 & $4\times900$& 1\arcsec$\times$20\arcsec & 47.0$^{d}$   & 600ZD & --  \\
5 & Q & 57546.532 & Keck1/LRIS & g & 30, 300 & -- & -- & -- & 22.9(1) \\ 
5 & Q & 57546.532 & Keck1/LRIS & R & 30, 300 & -- & -- & -- & 21.7(1) \\
5 & Q & 57546.532 & Keck1/LRIS & I & 150, 150 & -- & -- & -- & 21.6(1) \\
6 & A & 57661.265 & Keck1/LRIS & D560 & 900; 900 & 1\arcsec$\times$175\arcsec & 124.5   & 400/8500 \& 400/3400 & R=19.47(2)  \\
\enddata
\label{table:3}
\tablecomments{(\textit{a}) This refers to the optical activity state of the source (Q=quiescent, A=active). (\textit{b}) Imaging observations: the imaging filter is specified. Spectroscopic observations: dichroics are specified for LRIS, and the order-blocking filter is specified for DEIMOS. (\textit{c}) Spectroscopic observations: exposure sequences for the LRIS red and blue arms are separated by semi-colons. (\textit{d}) These slit PAs were chosen to place the slit perpendicular to the horizon at the time of observation, so as to minimize the loss of light due to atmospheric dispersion. (\textit{e}) Obtained under sub-photometric observing conditions, and not inclusive of systematic error of $\sim0.3$ magnitudes.}
\end{deluxetable*}

\subsection{The active counterpart}


\begin{figure}
\centering
\includegraphics[scale=0.5]{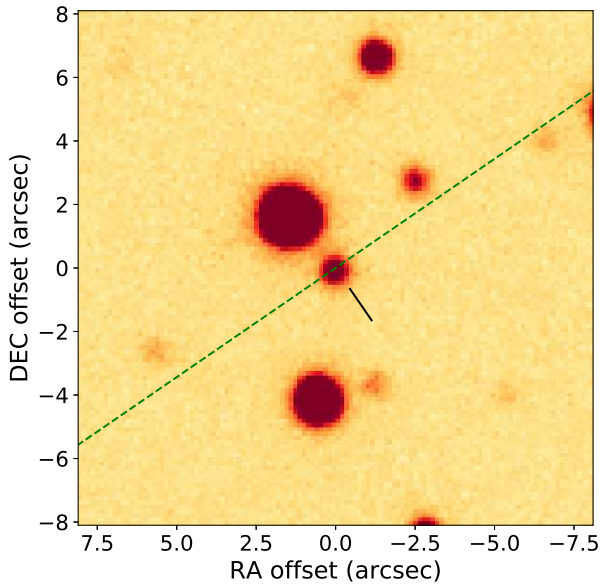}
\caption{I-band image of the quiescent optical counterpart to J1957 from the Epoch 5 observations (Table~\ref{table:3}). The counterpart is indicated by a black solid line. The image is centered on the position of J1957 (J2000): 19$^{\rm h}$56$^{\rm m}$39.11$^{\rm s}$, 03$^{\circ}$26\arcmin43.7\arcsec. The green dashed line indicates the slit PA ($124.5^{\circ}$) during the spectroscopic observations of Epochs 3 and 6.}
\label{fig:keck}
\end{figure}

I obtained a 900\,s spectrum of the optical counterpart with LRIS one day after the MJD\,57660 outburst reported by MAXI/GSC. 
Details of the instrument configuration are in Table~\ref{table:3} (Epoch 6). The slit position angle (PA) was chosen to avoid contamination from a nearby bright 
star; indeed, the object that confused the authors of ATels 8149 and 8197. The slit orientation is indicated in Figure~\ref{fig:keck}. The data were processed using standard 
procedures for this instrument. 
Bias subtraction using the overscan levels, flat fielding using dome-flat exposures, cosmic-ray rejection, optimal sky-line subtraction, 
and wavelength calibration using internal arc exposures corrected by sky-line positions were performed using D. Perley's {\it lpipe} 
software.\footnote{http://www.astro.caltech.edu/~dperley/programs/lpipe.html} Using the trace of the standard star BD$+$28$^{\circ}$4211, 
I then optimally extracted the spectra, and performed telluric absorption line and extinction corrections, flux calibration, and optimal addition 
of the spectra from the two arms of the spectrograph using the standard spectrum. The seeing was at 0.9\arcsec~full-width half-maximum (FWHM). The resulting spectrum is displayed in the top panel of Figure~\ref{fig:spec}. 

The spectrum is dominated by a smooth blue continuum between $3000-10000$\,\AA. This is consistent with the results reported by \citet{bkc+16} 
and \citet{mca+17} from a SALT spectrum of J1957 obtained approximately one day after these observations. 
Using the LRIS R-band filter profile, I obtain an AB magnitude of 
19.47(2), consistent with previous measurements of the optical source in outburst \citep{ryg15}. Besides the NaI~D doublet in absorption at $\sim5892$\,\AA, 
a selection of blue metallic emission lines are tentatively detected. These include HeII at 3203\,\AA~and 4686\,\AA, SII at 4072\,\AA, and SiII at 5041\,\AA, 
6347\,\AA, and 6371\,\AA. These lines were not detected in the SALT observations. 

\subsection{The quiescent counterpart}

The deepest imaging observations of the quiescent optical counterpart of J1957 were conducted on MJD\,57546 (Epoch 5, Table~\ref{table:3}). The observations 
were performed in the LRIS g, R, and I bands. The images were bias-subtracted using full-frame bias images, flat-fielded using dome-flat exposures, cleaned 
of cosmic ray hits, registered astrometrically against USNO B1.0 stars \citep{mlc+03}, and co-added using D. Perley's {\it lpipe} 
software. The seeing FWHM was 0.6\arcsec. The photometric zero-point in each filter was derived using same-night observations of SDSS photometric standard 
fields. I performed manual PSF photometry on the J1957 counterpart. I measured the following AB magnitudes:  $g = 22.9(1)$, $R = 21.7(1)$, $I = 21.6(1)$. The I-band image of 
J1957 is shown in Figure~\ref{fig:keck}. 
These measurements are consistent, to within errors, with those reported by \citet{bkc+16} and \citet{mca+17} of the quiescent counterpart. 
During the Epoch 2 imaging (Table~3), processing the data using identical techniques, I measured $g = 23.5(2)$ and $R = 22.3(2)$. Note however 
that although the Epoch 2 data had comparable exposure times, observing conditions were considerably worse and likely not photometric 
(seeing FWHM of 1\arcsec, thin high cirrus cloud). 

\begin{figure}
\centering
\includegraphics[angle=-90,scale=0.6]{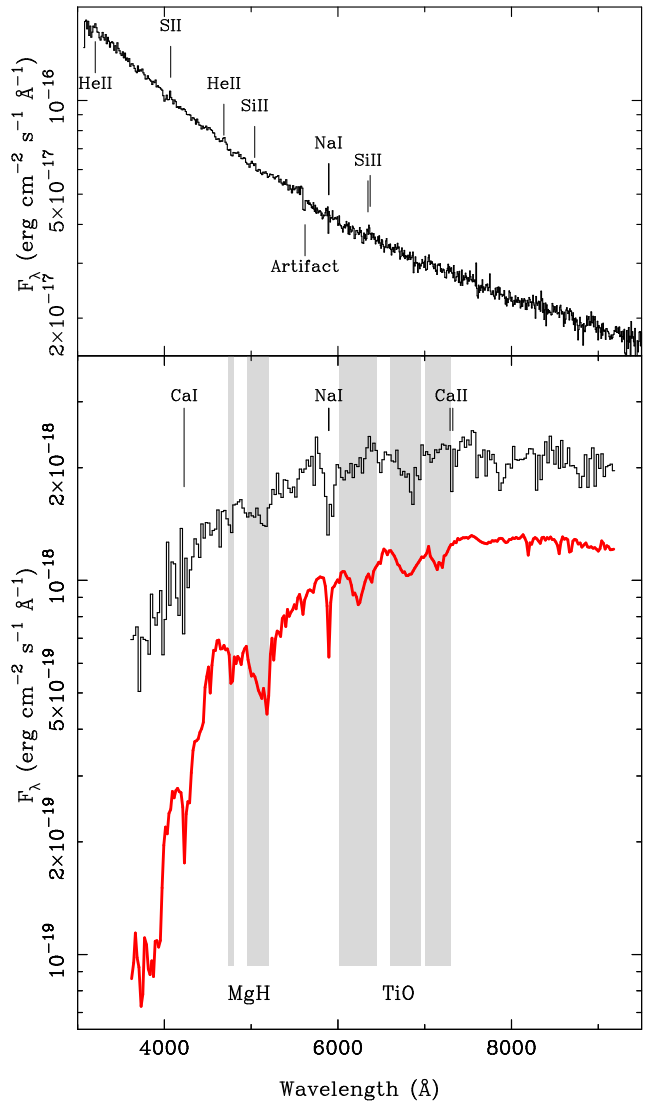}
\caption{Spectra of the optical counterpart to MAXI\,J1957+032 during an outburst (top -- Epoch\,6) and quiescence (bottom -- Epoch\,3). Relevant spectral lines and molecular absorption bands are indicated. The measured spectra are shown as binned data, and a template spectrum of a K7V dwarf from the Pickles (1998) spectral atlas is shown as a thick 
red trace in the bottom panel. The template is arbitrarily scaled in flux density for convenience of display. The quiescent spectrum shares numerous features with the K7V template, including the NaI and CaI narrow absorption features, as well as 
the broad MgH and TiO absorption bands. However, although this template is the best fit to the data from the Pickles atlas, it is apparent that the quiescent 
spectrum is bluer than the template. No interstellar extinction correction was applied to the data; the total Milky Way extinction along the J1957 sightline is 
expected to be $E(B-V)=0.11$.}
\label{fig:spec}
\end{figure}

The best spectroscopic observations of the quiescent counterpart were obtained with LRIS on MJD\,57519 (Epoch 3, Table~\ref{table:3}). I processed these 
data exactly as described above for the active counterpart; the standard star was Feige\,67, and the seeing FWHM was 0.6\arcsec. The resulting spectrum is 
shown in the bottom panel of Figure~\ref{fig:spec}. Again using the LRIS R-band filter profile, the spectrum is consistent with an AB magnitude of 22.7(5). However, 
as with the Epoch\,2 imaging data, thin high cirrus cloud present on this night may have made the conditions not suitable for accurate 
photometry. A red continuum was detected, 
along with a few absorption features (NaI D doublet at $\sim5892$\,\AA, potentially CaI at 4227\,\AA, and the CaII doublet at $\sim7310$\,\AA). A consistent but lower 
signal-to-noise ratio spectrum was measured with DEIMOS on MJD\,57543; the data were processed using the DEEP2 pipeline.\footnote{http://deep.ps.uci.edu/spec2d/} 

The broad absorption features at $\sim5000$\,\AA, and between $6000-7500$\,\AA, are consistent with MgH and TiO absorption bands in the atmosphere of a cool star, 
wherein the CaI and NaI D features are also expected.\footnote{Although the $5000-5100$\,\AA~band is in fact a blend of MgH and TiO absorption features, the extra absorption at the red end of the feature suggests that it is dominated by MgH absorption \citep{mkk43}.} 
I performed a quantitative comparison between the observed spectrum and the \citet{p98} library of stellar spectra by dividing the measured and template spectra by 
their continua (modeled as fourth-order polynomials), and measuring a $\chi^{2}$ statistic as a goodness of fit. The best-fitting spectrum was that of a K7V dwarf, 
with an M0V dwarf also providing an equally good match (factor of 1.01 worse $\chi^{2}$). The template continua were somewhat redder than the observed continuum, 
which may either be intrinsic to the source, or a result of inaccurate spectro-photometric calibration on the night. 
I show the scaled spectrum of the \citet{p98} K7V dwarf in Figure~\ref{fig:spec}. Giants of a similar spectral 
type (i.e., K7III, M0III) also provided a good fit (factor of 1.05 worse $\chi^{2}$), and it is thus difficult to distinguish between a cool dwarf and giant object spectroscopically. 

However, it is substantially more likely that the counterpart is a dwarf object than a giant. Late-K to early-M  
giants have absolute magnitudes brighter than $R=1$ \citep{c00}. J1957 is at Galactic coordinates $l=43.65^{\circ}$, $b=-12.83^{\circ}$, and I thus adopt a maximum 
distance of 20\,kpc (assuming that it lies within the Milky Way disk), implying a distance modulus of $<16.5$ magnitudes. If the counterpart were a giant, the R-band 
extinction along its sightline must be $A_{R}>4$ magnitudes for it to be consistent with our brightest apparent magnitude measurement. Let us now make the assumption that 
$A_{V}\approx A_{R}$ to estimate a lower limit on $E(B-V)$; this is conservative as typically $A_{V}\gtrsim A_{R}$. 
Then, the $A_{R}>4$ constraint implies $E(B-V)>1.3$ under typical Milky Way conditions. The expected Milky Way extinction along this sightline is 
characterized by $A_{R}=0.27$, $A_{V}=0.34$, and $E(B-V)=0.11$ \citep{sf11}. 

Using the relation between the equivalent width of the 
NaI D doublet (EW$_{\rm NaI\,D}$) and extinction identified by \citet{ppb12} for the Milky Way, I find EW$_{\rm NaI\,D}>1.7$\,\AA. Even if the conservative assumption were made 
that the blue continuum of the active counterpart to J1957 has no intrinsic NaI absorption, the measured equivalent width of the absorption feature is 
only $0.5(2)$\,\AA. This measurement is in fact consistent with the \citet{sf11} expectation for this sightline through the Milky Way. 
Furthermore, using the relation between $A_{V}$ and $n_{H}$ identified by \citet{go09}, an HI column of $n_{H}>8.8\times10^{21}$\,cm$^{-2}$ would be expected. 
Such a large column is not evident in the X-ray spectrum of J1957 \citep{cjm16,mca+17}. 

I therefore classify the quiescent counterpart to J1957 as a dwarf star of late-K or early-M spectral type. Note that these 
arguments are based on the assumption that the extinction does not significantly vary between active and quiescent states. Additionally, it is unlikely that the dwarf star is 
present at the position of J1957 by chance. The only optical image of J1957 in its active state was obtained by \citet{ryg15}, 
and their position is consistent with the centroid of the quiescent counterpart to within 0.3\arcsec, which is within the combined astrometric error 
of the \citet{ryg15} measurements and our astrometric solution (both are individually correct to 0.2\arcsec). The mean surface density of all detected objects in the 
Epoch\,5 $R$-band image is $\approx0.03$\,arcsec$^{-2}$, making a chance coincidence very unlikely ($\approx1/300$ for any object in our image). 
I hence proceed under the assumption that the late-K/early-M dwarf is physically associated with the J1957 system. 

\section{Interpretation}

\subsection{Empirical results}

Identification of the counterpart to J1957 permits an estimate of the distance to the system.  The fourth edition of Allen's Astrophysical Quantities \citep{c00} lists an absolute V-band magnitude of 8.8 for M0V stars\footnote{No listing is present for K7V stars. I therefore assume an M0V spectral type for the absolute-magnitude estimate.}, and a color of V$-$R=1.28. I estimated the uncertainty in the absolute magnitude by inspecting the publicly available HIPPARCOS catalog \citep{plk+97}. Following Figure 3 of \citet{plk+97}, I extracted the V-band magnitude and parallax measurements for all M0V stars with fractional errors in the parallax of $<10\%$, calculated absolute magnitudes and found a $1\sigma$ uncertainty of 0.8 magnitudes. An exact estimate of the absolute magnitudes of M0V stars from the parallax data is beyond the scope of this work, as such estimates are subject to significant biases \citep[e.g.,][]{ogs99}. I therefore adopt an absolute (color-corrected) R-band magnitude for the J1957 counterpart of 7.5(8). 

I use the Epoch~5 R-band magnitude of 21.7(1) for the J1957 counterpart to estimate its distance, as this epoch provided the most accurate photometric measurements. The HI column density of $n_{H}=1.7\times10^{21}$\,cm$^{-2}$  \citep{mca+17} implies a V-band extinction of $A_{V}=0.77(3)$ \citep{go09}, and thus $A_{R}=0.61(2)$ \citep{sf11}. This is somewhat larger than expected along this Milky Way sightline \citep{sf11}. The final distance estimate for the J1957 system is 5(2)\,kpc. To illustrate the sensitivity of this estimate to the assumed extinction, disregarding extinction would result in a distance of 6.9\,kpc, and an HI column of $n_{H}=4\times10^{21}$\,cm$^{-2}$ would imply a distance of 3.6\,kpc.

Having established a distance to J1957, the rest-frame properties of the 
source can be derived. These include the peak measured $0.5-10$\,keV X-ray luminosity, $L_{\rm XP}$, in the {\em Swift}/XRT windowed-timing data on MJD\,57660.69 \citep{mca+17}, the lowest firm $0.5-10$\,keV X-ray luminosity measurement, $L_{\rm XQ}$, from the {\it Swift}/XRT detection on MJD\,57193 (Table~\ref{table:xrt}), and the peak measured $R$-band luminosity, $L_{R}$. I list these properties in Table~4. I also include an estimate of the activity duty-cycle, $DC$, assuming (based on the MJD\,57153 flare) five days of activity per outburst. 

\begin{deluxetable}{cc}
\tabletypesize{\scriptsize}
\tablecaption{Properties of J1957.}
\tablewidth{0pt}
\tablehead{
\colhead{Quantity} & \colhead{Value} 
}
\startdata
$L_{\rm XP}$ & $10^{36.4(4)}$\,erg\,s$^{-1}$ \\
$L_{\rm XQ}$ & $10^{32.7(5)}$\,erg\,s$^{-1}$ \\
$L_{R}$ & $10^{32.1(4)}$\,erg\,s$^{-1}$ \\
$DC$ & $0.03$ \\
\enddata
\label{table:properties}
\end{deluxetable}

Despite large uncertainties in the parameter estimates, we now have a basis for comparing J1957 with known classes of Galactic X-ray 
transients. Although an exotic extragalactic origin for J1957 may be possible, if for example a background galaxy were confused with 
the quiescent stellar counterpart, I see no reason to consider such an unusual interpretation. Furthermore, the tentatively detected optical 
emission lines during the most recent J1957 outburst are not redshifted. 

J1957 shares many properties of low-mass X-ray binaries (LMXBs), with transient episodes of high accretion rates from a Roche-lobe overflowing star onto a compact object (NS or stellar-mass BH), known widely as X-ray novae \citep[for a review, see][]{ts96}. These systems exhibit outbursts of a few weeks to a few months, with duty cycles of $\sim10\%$, and with X-ray to optical flux ratios of $\sim10^{2}-10^{3}$. Although the non-detection of strong emission lines (in particular the Balmer series of hydrogen) during outburst is unusual \citep{vm95}, it is not exceedingly rare for a single-epoch observation \citep{dap+16}. The peak X-ray luminosity of J1957 is low with respect to the X-ray nova population; X-ray novae typically peak at $>10^{37}$\,erg\,s$^{-1}$, a factor of $\sim100$ greater than is observed for J1957. The outburst timescales, times between outbursts, and duty cycle of J1957 are also small with respect to the X-ray nova population. The faint-state X-ray luminosity, $L_{\rm XQ}$, is larger by a factor of a few hundred than has been observed for a BH LMXB \citep[e.g.,][]{br00}. 

As stated by \citet{mca+17}, the characteristic outburst duration, lightcurve shape, featureless optical spectrum, and duty cycle of J1957 are consistent with the sample of known AMXPs. The AMXP population \citep{pw12} includes objects such as NGC\,6440 X-2 \citep{aph+10}, with short outburst timescales of 3-5 days and recurrence times as short as 30 days, and a maximum 2-10\,keV luminosity of $1.7\times10^{36}$\,erg\,s$^{-1}$. In general, ultracompact AMXPs with orbital periods $P_{\rm orb}<80$\,min, and partially or fully degenerate companions exhibit outbursts as faint, short and often as J1957. Although \citet{mca+17} do not detect pulsations at frequencies in the range 0.1-281\,Hz from J1957 during its outbursts, several AMXPs spin more rapidly and do not always exhibit pulsations \citep{pw12}. 

However, although the luminosity I derive for J1957 based on the counterpart identification is consistent with AMXP outbursts, the main-sequence nature of the counterpart raises some questions. All AMXPs with main-sequence mass donors have exhibited thermonuclear Type-I X-ray bursts \citep{pw12}, while none have been observed in J1957. For a late-K/early-M dwarf companion, a lower main-sequence equation of state can be assumed, where the counterpart mass is given by $M_{c}/M_{\odot}\approx0.6$. This would be among the largest companion masses in AMXPs, comparable with {\em Swift}~J1749.4$-$2807 \citep{acp+11}. The variability I detect in the X-ray flux during the times between outbursts also requires an explanation. In particular, the source does not appear to fully fade after the MJD\,57153 outburst, but fades to a much greater extent following the MJD\,57302 and MJD\,57660 outbursts. Such variability in the putative quiescent state has only been observed in the AMXP IGR\,J18245$-$2452 \citep{lbh+14}, a system which transitions between a radio-pulsar state and AMXP-like outbursts. Below, I address these questions through theoretical considerations. 

\subsection{Theoretical insight}

\subsubsection{The accretion flow model}

I first characterize the ratio, $\Lambda$, between the luminosity of J1957 and the \citet{e16} luminosity. For ionized hydrogen plasma, 
this is given by 
\begin{equation}
\label{eqn:edd}
\Lambda = 8\times10^{-3}\frac{L_{X}}{10^{36}\,{\rm erg\,s^{-1}}}\left(\frac{M}{M_{\odot}}\right)^{-1},
\end{equation}
where $M$ is the mass of the compact object. Note that $\Lambda$ decreases for plasmas with significant metal fractions; a pure helium plasma, 
for example, would correspond to a factor of four decrease in $\Lambda$. $\Lambda$ also represents the ratio of the mass accretion rate, $\dot{M}$, to 
the Eddington-limiting mass accretion rate ($\dot{M}_{\rm Edd}$), irrespective of the radiative efficiency of the system. Hence, it appears that at 
the maximum of the X-ray lightcurve, J1957 accretes at $\dot{M}\approx0.014\dot{M}_{\rm Edd}$ for a $1.4M_{\odot}$ NS primary. I note that the 
correct X-ray luminosity to use in Equation~\ref{eqn:edd} when inferring the instantaneous ratio $\dot{M}/\dot{M}_{\rm Edd}$ is the 
instantaneous luminosity, not the time-averaged luminosity that I use below to derive the mean mass-loss rate from the secondary. I assume that the X-ray luminosity is approximately equivalent to the bolometric luminosity. 

The low Eddington ratio $\Lambda$ fully excludes a BH accretor in J1957. At low Eddington ratios, accretion in the standard cool, optically-thick accretion disk model \citep{ss73} is less likely to be stable than accretion in a hot, optically thin advection-dominated accretion flow \citep[ADAF;][]{ny95}. If the primary in J1957 is a BH, likely of mass $M\gtrsim6M_{\odot}$ \citep[i.e., as massive as, or more massive than the accretor in GRO\,J1655$-$40, which is the smallest mass confirmed for a stellar-mass BH;][]{ccm+16}, $\Lambda\lesssim3.3\times10^{-3}$. ADAFs are expected for BHs accreting at $\Lambda\lesssim\alpha^{2}$, where $\alpha\approx0.1$ is the accretion flow viscosity parameter \citep{ny95}. However, an ADAF configuration for the accretion implies that a much larger optical luminosity, of the same order of magnitude as the X-ray luminosity, would then be expected. This is not the case in J1957 \citep[see also][]{mca+17}. 

On the other hand, ADAFs are only expected for NSs accreting at $\Lambda\lesssim0.1\alpha^{2}\approx10^{-3}$. Therefore, disk accretion is likely for the bright states of J1957. This is not the case for the faint-state detections of J1957, wherein $\Lambda\approx2.9\times10^{-6}$. However, quiescent NS LMXBs can emit X-rays through a variety of mechanisms, including thermal emission from the NS surface, and non-thermal magnetospheric emission \citep[e.g.,][]{csk08,lbh+14}. Chromospheric activity in the main-sequence counterpart is unlikely due to the lack of the expected optical emission-line features during quiescence (e.g., CaII H and K, Balmer-series hydrogen). 

\subsubsection{The mass accretion rate}

I now consider the implications of the time-averaged mass accretion rate of the J1957 primary. Assuming a typical outburst bolometric 
luminosity of $10^{36}$\,erg\,s$^{-1}$, a duty cycle of 0.03, a typical quiescent luminosity of $10^{32}$\,erg\,s$^{-1}$, and that the 
radiation is isotropic, the time-averaged mass accretion rate is 
\begin{equation}
\dot{M} \approx 5\times10^{-13}\left(\frac{1}{\eta}\right)\,M_{\odot}\,{\rm yr^{-1}}.
\end{equation}
Here, $\eta$ is the time-averaged (and luminosity-weighted) efficiency of the conversion between accreted mass and radiated energy. This, however, is clearly a lower limit. 
An upper limit under the same assumptions can be derived 
by assuming that only the gravitational potential energy of the accretion flow once at the compact object surface is converted to radiation. 
Then, for an NS accretor, 
\begin{equation}
\dot{M} \lesssim 3\times10^{-12}\frac{R}{\rm 10\,km}\left(\frac{M}{1.4M_{\odot}}\right)^{-1}\,M_{\odot}\,{\rm yr^{-1}},
\end{equation}
where $R$ is the NS radius. 

I compare these estimates with the expected Roche-lobe overflow mass-loss rate for a late-K/early-M dwarf star, following arguments 
by \citet{k88}. Assuming a lower main-sequence equation of state for the star as before, the stellar radius is given by $R_{c}/R_{\odot}=M_{c}/M_{\odot}\approx0.6$. Then, 
equating the stellar radius with the size of its Roche lobe, and assuming a circular Keplerian orbit, the orbital period is 
$P_{\rm orb}\approx6$\,h \citep[][Equation~4]{k88}. The minimum Roche-lobe overflow mass-loss rate can then be estimated 
by ascribing the shrinking of the Roche lobe to orbital decay driven only by gravitational radiation. From Equation~14 of 
\citet{k88}, this is given by $\dot{M}\gtrsim5\times10^{-11}\,M_{\odot}\,{\rm yr^{-1}}$. Therefore, a standard Roche-lobe 
overflow configuration for J1957 is unlikely for an NS accretor, if the mass donor is the late-K/early-M dwarf star. This result is robust to a factor of ten uncertainty in the luminosity. 

Other mass-loss mechanisms for the late-K/early-M dwarf are also unlikely to feed an NS primary at the inferred accretion rate. Based on analyses of metal-rich 
white dwarf atmospheres where M-dwarf companions are present, \citet{d06} suggests wind mass-loss rates of 
$10^{-16}-10^{-14}\,M_{\odot}\,{\rm yr^{-1}}$ for M-dwarfs. Through the analysis of ``astrospheric'' Ly~$\alpha$ absorption features from the 
collision regions between stellar winds and the interstellar medium, dwarf main-sequence stars are found to have comparable mass-loss rates 
to the Sun \citep[i.e., $\approx10^{-14}\,M_{\odot}\,{\rm yr^{-1}}$;][]{wmz+05}. Somewhat higher mass-loss rates may be expected if X-ray irradiation of the 
secondary by an NS primary significantly affects the surface temperature. However, assuming a 3\,h orbit for the system, and a 
quiescent luminosity of J1957 due to thermal emission from the surface of an NS primary, $\approx10^{31}$\,erg\,s$^{-1}$ will be incident 
on the companion. Thermal equilibrium suggests that this would raise the surface temperature by $\sim10^{3}$\,K. A detailed model for the 
effects of X-ray irradiation on companions in LMXBs was presented by \citet{hkl86}, who showed that the instantaneous accretion rate cannot 
be expected to be steady due to irradiation effects for binary periods $\lesssim3$\,h. However, a time-averaged accretion rate of $>10^{-13}\,M_{\odot}\,{\rm yr^{-1}}$ is unlikely to be sustained through this mechanism. 

\subsection{An unidentified mass donor?} 

It is possible that the quiescent counterpart of J1957 is not in fact the star from which the compact object is accreting. This could be the 
case in two scenarios: (\textit{a}) if the late-K/early-M dwarf is at the sky-position of J1957 by chance, or (\textit{b}) if the dwarf is part of the J1957 
system, but forms a wide triple with a compact interacting binary. I have already rejected possibility (\textit{a}), and I therefore adopt possibility (\textit{b}). 

In this scenario, the most 
compelling candidate for the donor object would be a degenerate star, likely low-mass, in a tight orbit with the primary. Two ways of achieving the required 
low mass-transfer rates have been suggested for low-luminosity transients like J1957. 
First, once an LMXB has evolved to an orbital period of $\sim80$\,min where the donor reaches a mass of 
$\lesssim0.1M_{\odot}$ and becomes partially degenerate, the mass transfer rate is slowed to 
$\lesssim10^{-11}\,M_{\odot}\,{\rm yr^{-1}}$ 
\citep{k00}. The orbital period increases after this point, as further mass loss causes the donor to expand. 
This state is stable for many Gyr, and such systems are expected to be transients owing to the thermal-viscous disk 
instability and the low mass-transfer rates, and dominated by NS accretors because of the predominance of NSs in all LMXBs (transient 
or persistent). Second, similar behavior is expected for ultracompact LMXBs that are the NS analogs of AM CVn systems \citep{hl16}, 
with the main distinguishing feature from the first possibility being a somewhat higher-mass degenerate companion, and an orbital period 
below $\sim80$\,min. 

Objects in both the configurations mentioned above are expected to be common in the Milky Way, and are the leading candidates to explain 
X-ray transients with low inferred mass-transfer rates where no optical counterparts are identified \citep{mpb+05}. An attractive feature of 
both models is the expected lack of Balmer-series hydrogen or HeI emission lines in the spectrum of J1957 during outburst. These species are 
generally observed in LMXB outbursts \citep[e.g.,][]{vm95}, and may be expected to be present if the donor star in J1957 were a late-K/early-M 
dwarf as assumed above \citep[although not necessarily for short-period binaries;][]{dap+16}. The outburst characteristics of J1957 are in fact quite similar to 
those of the transient ultracompact LMXB candidate 
1RXS\,J180408.9$-$34205 \citep{bdc+16,dap+16}. Indeed, the INTEGRAL hard X-ray detection of J1957 during the MJD\,57153 outburst 
that possibly preceded the MAXI detection at softer energies may have represented a hard state of the system that transitioned to a soft 
state around the epoch of the MAXI observations, in analogy with the outburst of 1RXS\,J180408.9$-$34205. Further, the spectrum of 
1RXS\,J180408.9$-$34205 during outburst is similarly featureless to that of J1957, with the tentative detection of the 4686\,\AA~HeII line in 
emission. 

A second attractive feature of these models is the expectation of no thermonuclear bursts. For a pure helium or carbon-oxygen white dwarf donor, no hydrogen is expected to be present in the accreting material. Additionally, the AMXP IGR\,00291$+$5934 with a brown-dwarf companion has also not displayed Type-I bursts, although this is not the case for two other AMXPs with brown-dwarf companions \citep{pw12}. 

A white-dwarf or brown dwarf donor in the J1957 system would not be detected in my observations during quiescence. Typical WDs are approximately 
two to four magnitudes less luminous than the luminosity I assume for the early-K/late-M dwarf star, and brown dwarfs are several magnitudes fainter still. 

\subsection{Summary of interpretation}

I propose that J1957 consists of an NS in a compact ($P_{\rm orb}\lesssim2$\,h) mass-transferring binary system, with a hydrogen- and possibly helium-poor donor. A carbon-oxygen white dwarf is preferred as the mass donor from a comparison with other accreting NSs. The optical counterpart identified during quiescence is then likely to be in a wide orbit around the binary; a few-hundred day orbital period is preferred due to stability considerations \citep{ma01}. 

Approximately 20\% of multiple-star systems are found in triples \citep{et08}, and two NSs in triple systems are known \citep{rsa+14,gbg+89}. One of these systems is the LMXB 4U\,2129$+$47, which consists of an NS accreting from a main-sequence dwarf star in a 4.27\,h orbit, with an additional F7V star in a $>175$\,day orbit \citep{gbg+89,btg+08}. As the orbital period of the interacting binary is smaller than a day, the binary will likely shrink through the loss of energy and angular momentum to magnetic-dipole and gravitational radiation to form an ultracompact LMXB with a degenerate donor \citep{ps88,vnv+13}. Such a system, but with a lower-mass tertiary, would be analogous to the scenario that I favor for J1957. 

The formation mechanisms and rates of NS triple systems such as the one proposed for J1957 have been investigated by \citet{fbw+11} and \citet{pvv+11}. Accreting NSs in hierarchical triples are the leading candidate progenitors for the formation of millisecond pulsars observed in wide binaries with main-sequence companions, including  PSR\,J1903$+$0327 \citep{crl+08} and PSR\,J1024$-$0719 \citep{bjs+16,kkn+16}. These systems are posited to have begun as hierarchical main-sequence stellar triples, where the most massive component underwent a supernova and formed an NS. The pre-supernova common envelope evolution phase is expected to have shrunk the initial orbits, such that after the supernova the inner main-sequence star can fill and overflow its Roche lobe. If the initial orbit upon mass transfer is sufficiently large, further evolution of the mass donor will cause the orbit to grow, and dynamical instabilities will eventually lead to its ejection from the system. On the other hand, if the initial orbit were sufficiently small, the interacting-binary orbit can shrink as described above, and the mass donor would evolve off the main sequence, forming a system like J1957. The mass donor may then be ablated by the NS once it has gained sufficient angular momentum to form a millisecond pulsar.

\section{Summary and future prospects}

MAXI\,J1957$+$032 (J1957) is an unusual Galactic X-ray transient. It flares by a factor of $\gtrsim10^{4}$ once every few-hundred days, with each flare typically 
lasting just $\approx5$ days. I have identified its quiescent optical counterpart to be a late-K/early-M dwarf star, and thus establish the distance to the 
system to be $5(2)$\,kpc. An optical spectrum obtained during a flare reveals a blue continuum with weak metallic emission lines. 

The distance to J1957 implies a peak $0.5-10$\,keV luminosity of $10^{36.4(4)}$\,erg\,s$^{-1}$. Faint X-ray emission with a luminosity of $10^{32.7(5)}$\,erg\,s$^{-1}$ is detected on some epochs between flares, with upper limits that are a few times lower on other epochs. As also found by \citet{mca+17}, the outburst behavior of J1957 is most consistent with the accreting millisecond X-ray pulsar (AMXP) sample. However, the low time-averaged mass accretion rate that I infer of approximately $10^{-12}\,M_{\odot}$\,yr$^{-1}$ is difficult to reconcile with Roche-lobe overflow of the late-K/early-M dwarf counterpart, if this counterpart is indeed the mass donor. Further pieces of circumstantial evidence argue against a main-sequence mass donor: no Balmer-series hydrogen or HeI emission lines are present in the optical spectrum of J1957 in its flaring state, and no thermonuclear Type-I X-ray bursts have been detected. Therefore, I hypothesize that the mass donor is not detected in the optical observations I present, but is instead a degenerate hydrogen- and possibly helium-poor star in a tight binary system with the NS. A comparison with the AMXP sample \citep{pw12} suggests an ultracompact ($P_{\rm orb}<80$\,min) binary with a carbon-oxygen white dwarf mass donor as the preferred candidate. In this case, J1957 is likely to be a triple system, with the observed main-sequence counterpart in a wide orbit around the mass-transferring binary. Such systems are predicted to exist along the leading formation channel for millisecond pulsars observed in wide binaries with main-sequence companions \citep{fbw+11,pvv+11}. 

To make progress towards a better observational understanding of J1957, a measurement of the orbital period of the system, and a direct determination of the nature 
of the mass donor, are crucial. The most valuable observations would be multiple epochs of optical spectroscopy during an outburst. The emission lines that were observed in my Epoch~6 data, if indeed present, are clearly comparable in flux density to the hot thermal continuum. Their equivalent widths would increase if the accreting gas is allowed to cool, which is highly likely to occur as the system returns to quiescence. Thus, late-time optical spectroscopy of an outburst can yield more information on the composition of the accretion flow. The presence of hydrogen is expected in the case of a late-K/early-M dwarf donor. If 
the gas temperature during my observations was $\gtrsim2\times10^{4}$\,K, under reasonable pressures it is unlikely that  Balmer-series hydrogen lines could have been observed. However, these lines would be expected at later times. Moderate-resolution ($\sim3$\,\AA~FWHM) time-resolved spectroscopy would also provide sensitivity to any orbital period $\lesssim10$\,h through Doppler shifts of spectral lines. If the late-K/early-M dwarf star is the mass donor, an orbital period of more than a few hours may be expected. A somewhat shorter orbital period would be observed if the system is an ultracompact LMXB in a triple system with the main-sequence star.

\acknowledgments

I thank S. Kulkarni for introducing me to this object, N. Blagorodnova, Y. Cao, G. Duggan, and R. Lunnan for 
assistance with the optical observations, and S. Phinney and T. Maccarone for useful discussions.  The data presented herein were obtained at the W. M. Keck Observatory, which is operated as a scientific partnership among the California Institute of Technology, the University of California and the National Aeronautics and Space Administration. The Observatory was made possible by the generous financial support of the W. M. Keck Foundation. I recognize and acknowledge the very significant cultural role and reverence that the summit of Maunakea has always had within the indigenous Hawaiian community. I acknowledge the use of public data from the Swift data archive. This research has made use of the VizieR catalogue access tool, CDS,
 Strasbourg, France. 

\facility{Keck:I (LRIS), Keck:II (DEIMOS), MAXI, Swift}.


\begin{thebibliography}{}

\bibitem[Altamirano et al.(2010)]{aph+10} Altamirano, D., Patruno, A., Heinke, C.~O., et al.\ 2010, ApJ, 712, L58 
\bibitem[Altamirano et al.(2011)]{acp+11} Altamirano, D., Cavecchi, Y., Patruno, A., et al.\ 2011, ApJ, 727, L18 
\bibitem[Baglio et al.(2016)]{bdc+16} Baglio, M.~C., D'Avanzo, P., Campana, S., et al.\ 2016, A\&A, 587, A102
\bibitem[Bassa et al.(2016)]{bjs+16} Bassa, C.~G., Janssen, G.~H., Stappers, B.~W., et al.\ 2016, MNRAS, 460, 2207 
\bibitem[Bildsten \& Rutledge(2000)]{br00} Bildsten, L., \& Rutledge, R.~E.\ 2000, ApJ, 541, 908
\bibitem[Bothwell et al.(2008)]{btg+08} Bothwell, M.~S., Torres, M.~A.~P., Garcia, M.~R., \& Charles, P.~A.\ 2008, A\&A, 485, 773 
\bibitem[Buckley et al.(2016)]{bkc+16} Buckley, D.~A.~H., Kotze, M.~A., Charles, P.~A., et al.\ 2016, The Astronomer's Telegram, 9649
\bibitem[Burrows et al.(2005)]{bhn+05} Burrows, D.~N., Hill, J.~E., Nousek, J.~A., et al.\ 2005, SSR, 120, 165
\bibitem[Campana et al.(2008)]{csk08} Campana, S., Stella, L., \& Kennea, J.~A.\ 2008, \apjl, 684, L99 
\bibitem[Chakrabarty et al.(2016)]{cjm16} Chakrabarty, D., Jonker, P.~G., \& Markwardt, C.~B.\ 2016, The Astronomer's Telegram, 9591
\bibitem[Champion et al.(2008)]{crl+08} Champion, D.~J., Ransom, S.~M., Lazarus, P., et al.\ 2008, Science, 320, 1309
\bibitem[Cherepashchuk et al.(2015)]{cml+15} Cherepashchuk, A.~M., Molkov, S.~V., Lutovinov, A.~A., \& Postnov, K.~A.\ 2015, The Astronomer's Telegram, 7506
\bibitem[Corral-Santana et al.(2016)]{ccm+16} Corral-Santana, J.~M., Casares, J., Mu{\~n}oz-Darias, T., et al.\ 2016, A\&A, 587, A61
\bibitem[Cox(2000)]{c00} Cox, A.~N.\ 2000, Allen's Astrophysical Quantities,  fourth edition, AIP Press (New York)
\bibitem[Debes(2006)]{d06} Debes, J.~H.\ 2006, ApJ, 652, 636
\bibitem[Degenaar et al.(2016)]{dap+16} Degenaar, N., Altamirano, D., Parker, M., et al.\ 2016, MNRAS, 461, 4049
\bibitem[Eddington(1916)]{e16} Eddington, A.~S.\ 1916, MNRAS, 77, 16
\bibitem[Eggleton \& Tokovinin(2008)]{et08} Eggleton, P.~P., \& Tokovinin, A.~A.\ 2008, MNRAS, 389, 869
\bibitem[Faber et al.(2003)]{fpk+03} Faber, S.~M., Phillips, A.~C., Kibrick, R.~I., et al.\ 2003, Proc.~SPIE, 4841, 1657
\bibitem[Freire et al.(2011)]{fbw+11} Freire, P.~C.~C., Bassa, C.~G., Wex, N., et al.\ 2011, MNRAS, 412, 2763
\bibitem[Garcia et al.(1989)]{gbg+89} Garcia, M.~R., Bailyn, C.~D., Grindlay, J.~E., \& Molnar, L.~A.\ 1989, ApJ, 341, L75
\bibitem[G{\"u}ver \& {\"O}zel(2009)]{go09} G{\"u}ver, T., \& {\"O}zel, F.\ 2009, MNRAS, 400, 2050
\bibitem[Hameury et al.(1986)]{hkl86} Hameury, J.~M., King, A.~R., \& Lasota, J.~P.\ 1986, A\&A, 162, 71
\bibitem[Hameury \& Lasota(2016)]{hl16} Hameury, J.-M., \& Lasota, J.-P.\ 2016, A\&A, 594, A87
\bibitem[Kaplan et al.(2016)]{kkn+16} Kaplan, D.~L., Kupfer, T., Nice, D.~J., et al.\ 2016, ApJ, 826, 86
\bibitem[King(1988)]{k88} King, A.~R.\ 1988, QJRAS, 29, 1 
\bibitem[King(2000)]{k00} King, A.~R.\ 2000, MNRAS, 315, L33
\bibitem[Linares et al.(2014)]{lbh+14} Linares, M., Bahramian, A., Heinke, C., et al.\ 2014, MNRAS, 438, 251
\bibitem[Mardling \& Aarseth(2001)]{ma01} Mardling, R.~A., \& Aarseth, S.~J.\ 2001, MNRAS, 321, 398
\bibitem[Mata S{\'a}nchez et al.(2017)]{mca+17} Mata S{\'a}nchez, D., Charles, P.~A., Armas Padilla, M., et al.\ 2017, arXiv:1701.03381
\bibitem[Matsuoka et al.(2009)]{mku+09} Matsuoka, M., Kawasaki, K., Ueno, S., et al.\ 2009, PASJ, 61, 999 
\bibitem[Morgan et al.(1943)]{mkk43} Morgan, W.~W., Keenan, P.~C., \& Kellman, E.\ 1943, Chicago, Ill., The University of Chicago press [1943]
\bibitem[Molkov et al.(2015)]{mlp+15} Molkov, S.~V., Lutovinov, A.~A., Postnov, K.~A., \& Cherepashchuk, A.~M.\ 2015, The Astronomer's Telegram, 7520
\bibitem[Monet et al.(2003)]{mlc+03} Monet, D.~G., Levine, S.~E., Canzian, B., et al.\ 2003, AJ, 125, 984
\bibitem[Muno et al.(2005)]{mpb+05} Muno, M.~P., Pfahl, E., Baganoff, F.~K., et al.\ 2005, ApJ, 622, L113
\bibitem[Narayan \& Yi(1995)]{ny95} Narayan, R., \& Yi, I.\ 1995, ApJ, 452, 710
\bibitem[Negoro et al.(2010)]{nmo+10} Negoro, H., Miyoshi, S., Ozawa, H., et al.\ 2010, Astronomical Data Analysis Software and Systems XIX, 434, 127 
\bibitem[Negoro et al.(2015)]{nsm+15} Negoro, H., Serino, M., Mihara, T., et al.\ 2015, The Astronomer's Telegram, 7504
\bibitem[Negoro et al.(2016)]{nst+16} Negoro, H., Sasaki, R., Tomida, H., et al.\ 2016, The Astronomer's Telegram, 9565
\bibitem[Oke et al.(1995)]{occ+95} Oke, J.~B., Cohen, J.~G., Carr, M., et al.\ 1995, PASP, 107, 375
\bibitem[Oudmaijer et al.(1999)]{ogs99} Oudmaijer, R., Groenewegen, M.~A.~T., \& Schrijver, H.\ 1999, A\&A, 341, L55 
\bibitem[Patruno \& Watts(2012)]{pw12} Patruno, A., \& Watts, A.~L.\ 2012, arXiv:1206.2727
\bibitem[Perryman et al.(1997)]{plk+97} Perryman, M.~A.~C., Lindegren, L., Kovalevsky, J., et al.\ 1997, A\&A, 323, L49 
\bibitem[Pickles(1998)]{p98} Pickles, A.~J.\ 1998, PASP, 110, 863
\bibitem[Portegies Zwart et al.(2011)]{pvv+11} Portegies Zwart, S., van den Heuvel, E.~P.~J., van Leeuwen, J., \& Nelemans, G.\ 2011, ApJ, 734, 55 
\bibitem[Poznanski et al.(2012)]{ppb12} Poznanski, D., Prochaska, J.~X., \& Bloom, J.~S.\ 2012, MNRAS, 426, 1465
\bibitem[Pylyser \& Savonije(1988)]{ps88} Pylyser, E., \& Savonije, G.~J.\ 1988, A\&A, 191, 57
\bibitem[Ransom et al.(2014)]{rsa+14} Ransom, S.~M., Stairs, I.~H., Archibald, A.~M., et al.\ 2014, Nature, 505, 520
\bibitem[Rau et al.(2015)]{ryg15} Rau, A., Yates, R., \& Greiner, J.\ 2015, The Astronomer's Telegram, 7524
\bibitem[Roming et al.(2005)]{rkm+05} Roming, P.~W.~A., Kennedy, T.~E., Mason, K.~O., et al.\ 2005, SSR, 120, 95
\bibitem[Schlafly \& Finkbeiner(2011)]{sf11} Schlafly, E.~F., \& Finkbeiner, D.~P.\ 2011, ApJ, 737, 103
\bibitem[Shakura \& Sunyaev(1973)]{ss73} Shakura, N.~I., \& Sunyaev, R.~A.\ 1973, A\&A, 24, 337
\bibitem[Sugimoto et al.(2015)]{sns+15} Sugimoto, J., Negoro, H., Sugizaki, M., et al.\ 2015, The Astronomer's Telegram, 8143
\bibitem[Tanaka \& Shibazaki(1996)]{ts96} Tanaka, Y., \& Shibazaki, N.\ 1996, ARA\&A, 34, 607
\bibitem[Tanaka et al.(2016)]{tnu+16} Tanaka, K., Negoro, H., Ueno, S., et al.\ 2016, The Astronomer's Telegram, 8529
\bibitem[van Haaften et al.(2013)]{vnv+13} van Haaften, L.~M., Nelemans, G., Voss, R., et al.\ 2013, A\&A, 552, A69 
\bibitem[van Paradijs \& McClintock(1995)]{vm95} van Paradijs, J., \& McClintock, J.~E.\ 1995, X-ray Binaries, 58
\bibitem[Wood et al.(2005)]{wmz+05} Wood, B.~E., M{\"u}ller, H.-R., Zank, G.~P., Linsky, J.~L., \& Redfield, S.\ 2005, ApJ, 628, L143

\end{thebibliography}
\end{document}